\documentclass[aps,showpacs,twocolumn]{revtex4}

\usepackage{graphicx}
\usepackage{booktabs}
\usepackage{amssymb,bm,mathrsfs,bbm,amscd}
\usepackage[tbtags]{amsmath}
\usepackage{lastpage}
\usepackage{CJK}

\begin{document}


\title{Dynamical study of the light scalar mesons below 1 GeV in a flux-tube
model\thanks{Supported by National Natural Science Foundation of
China under Contract No.: 11047140, 11175088 and 11035006 and the
Ph.D Program Funds of Chongqing Jiaotong University. }}

\author{DENG Cheng-Rong$^{1}$, PING Jia-Lun$^{2;1)}$\email{jlping@njnu.edu.cn
(corresponding author)}, ZHOU Ping$^{1}$, WANG
Fan$^3$}

\address{%
$^1$School of Mathematics and Physics, Chongqing
Jiaotong University, Chongqing 400074, China \\
$^2$ School of Physical Science and Technology, Nanjing Normal University, Nanjing
210097, China \\
$^3$Department of Physics, Nanjing University,
Nanjing 210093, P.R. China}

\begin{abstract}
The light scalar mesons below 1 GeV as tetraquark systems are
studied in the framework of the flux-tube model. Comparative
studies indicate that a multi-body confinement instead of the
additive two-body confinement sould be used in a multiquark
system. The mesons $\sigma$ and $\kappa$ could be well
accommodated in the diquark-antidiquark tetraquark picture and
could be colour confinement resonances. The mesons $a_0(980) $ and
$f_0(980)$ are not described as $K\bar{K}$ molecular states and
$ns\bar{n}\bar{s}$ diquark-antidiquark states. However, the mass
of the first radial excited state of the diquark-antidiquark state
$nn\bar{n}\bar{n}$ is 1019 MeV, which is close to the experimental
data of the $f_0(980)$.
\end{abstract}



\maketitle


\section{Introduction}

The charged $\kappa$, a scalar meson, was recently observed by BES
Collaboration~\cite{bes2}. The Breit-Wigner mass and the decay
width are obtained to be $826\pm49^{+49}_{-34}$ MeV and
$449{\pm}156^{+144}_{-81}$ MeV, and the pole position is
determined to be $(764\pm63^{+71}_{-54})-i(306\pm 149^{+143}_{-85}
)$ MeV/c$^2$. They are in good agreement with those of the neutral
$\kappa$: the mass and the decay width are $878\pm 23^{+64}_{-55}$
MeV and $499\pm 53^{+55}_{-87}$ MeV, respectively, observed by the
BES and other collaborations~\cite{neutralk}.

The understanding of scalar mesons, which have the same quantum
numbers as the vacuum, is a crucial problem in low-energy quantum
chromodynamics (QCD) since they could shed light on the chiral
symmetry breaking mechanism and presumably also on confinement in
QCD. Although many properties of scalar mesons have been studied
for decades, it is still a puzzle for the understanding of the
internal structure of the scalar mesons. Their masses do not fit
into the quark model predictions~\cite{jaffe3,Amsler}. The flavour
structures of these light scalar mesons below 1 GeV, $a_0(980)$,
$f_0(980)$, $\sigma$ and $\kappa$, are still an open question. In
the $q\bar{q}$ configuration, the $p$-wave relative motion between
$q$ and $\bar{q}$ has to be invoked to account for the spin and
parity of the scalar mesons. This leads to much higher masses for
them. Another possible configuration for scalar mesons is a
tetraquark state. In the tetraquark configuration, the light
scalar mesons could be classified into an SU(3) flavour nonet if
the diquark picture is used~\cite{jaffe1,jaffe2,alford,maiani}.
Their quark contents can be expressed as
\begin{eqnarray}
\sigma&=&[ud][\bar{u}\bar{d}],f_0^0=\frac{[su][\bar{s}\bar{u}]+[sd]
[\bar{s}\bar{d}]}{\sqrt{2}};\nonumber\\
\kappa^+&=&[ud][\bar{d}\bar{s}],\bar{\kappa^+}=[ds][\bar{u}\bar{d}],
\kappa^0=[ud][\bar{u}\bar{s}],\bar{\kappa^0}=[us][\bar{u}\bar{d}];\nonumber\\
a_0^+&=&[su][\bar{s}\bar{d}],a_0^0=\frac{[su][\bar{s}\bar{d}]-[sd]
[\bar{s}\bar{u}]}{\sqrt{2}},a_0^-=[sd][\bar{s}\bar{u}].\nonumber
\end{eqnarray}
Jaffe {\em et al.} interpreted light scalar mesons as tetraquark
states with all the relative orbital angular momenta assumed to be
zero~\cite{jaffe1,jaffe2,alford,maiani,Pelaez1,Pelaez2,thooft}.
Weinstein {\em et al.} described light scalar mesons as hadronic
molecular states due to strong meson-meson
interaction~\cite{weinstein0,giacosa,branz1,branz2,Janssen,Kaminski,Oller1,Oller2}.
The properties of some of these light scalar mesons were also
studied in the $q\bar{q}$ picture~\cite{van,celenza,Umekawa}. The
spectrum of light scalar mesons below 1.0 GeV were studied in the
$q\bar{q}$ picture by including instanton interaction~\cite{Dai}.
Bhavyashri {\em et al.} studied the instanton-induced interaction
in light meson spectrum on the basis of the phenomenological
harmonic models for quarks~\cite{Bhavyashri}. Vijande {\em et al.}
studied the scalar mesons in terms of the mixing of a chiral nonet
of tetraquarks with conventional $q\bar{q}$
states~\cite{Vijande,Napsuciale}.

A multi-quark state is quite different from ordinary hadrons
($q\bar{q}$ meson and $qqq$ baryon) because the multi-quark state
has more colour structures than those of ordinary hadrons. The
colour structures of a multi-quark state are no longer trivial and
the properties of the multi-quark states may be sensitive to the
hidden colour structure. A tetraquark state, if its existence is
confirmed, may provide important information about the low-energy
QCD interaction which is absent from the ordinary hadrons. Some
authors had studied the tetraquark system with the three-body
$qq\bar{q}$ and $q\bar{q}\bar{q}$
interaction~\cite{Dmitrasinovic,Pepin}. The exotic hadrons were
also studied as multiquark states in the flux tube model in our
previous work~\cite{deng,ping}. These studies suggest that the
multi-body confinement, instead of the additive two-body
confinement, might be more suitable in the quark model study of
multiquark states. The newly updated experimental data might shed
more light on the possibility of the existence of tetraquark
states and QCD interaction for multi-quark states.

The aim of this paper is to study the properties of scalar mesons
below 1 GeV in the flux tube model with multi-body confinement
potential. The powerful method for few-body systems with high
precision, Gaussian expansion method (GEM)~\cite{GEM}, is used
here. The paper is organized as follows: in Sec. 2, the flux-tube
model with multi-body interaction is introduced. A brief
introduction of GEM and the construction of the wave functions of
tetra-quark states are given in Sec. 3. The numerical results and
discussions are presented in Sec. 4 and a brief summary is given
in the last section.

\section{Quark model and multi-body confinement potential}

Long-term studies in the past several decades on hadrons indicate
that ordinary hadrons ($q\bar{q}$ meson and $qqq$ baryon) can be
well described by QCD-inspired quark models. Low energy QCD
phenomena are dominated by two well known quark correlations:
confinement and chiral symmetry breaking. The perturbative,
effective one gluon exchange, properties of QCD should also be
included. Hence, the main ingredients of the quark model are:
constituent quarks with a few hundred MeV effective mass,
phenomenological confinement potential, effective Goldstone bosons
and one-gluon exchanged between these constituent quarks.

For ordinary hadrons, the colour structures of them are unique and
trivial, naive models based on two-body colour confinement
interactions proportional to the colour charges
$\mathbf{\lambda}_i\cdot\mathbf{\lambda}_j$ can describe the
properties of ordinary hadrons well. However, the structures of a
multiquark state are abundant~\cite{deng,ping,ww}, which include
important QCD information that is absent from ordinary hadrons.
There is not any theoretical reason to extend directly the
two-body confinement in naive quark model to a multi-quark system.
Furthermore, the direct application of the two body confinement to
multi-quark system induces many serious problems, such as
anti-confinement~\cite{Dmitrasinovic} and colour Van der Waals
force. Much theoretical work has been done to try to amend those
serious drawbacks. The string flip model for multi-quark system
was proposed by M. Oka to avoid the pathological Van der Waals
force~\cite{oka1,oka2}. Three quark confinement is explored by
introducing strings which connect quarks according to a certain
configuration rule.

Recent lattice QCD studies~\cite{lattice,lattice0,latt1} show that
the confinement of multi-quark states is a multi-body interaction
and is proportional to the minimum of the total length of strings
which connect the quarks to form a multiquark state. Based on
these studies, a naive flux-tube or string
model~\cite{deng,ping,ww} with multi-body confinement has been
proposed for multiquark systems. The harmonic interaction
approximation, i.e., the total length of strings is replaced by
the sum of the square of string lengths, is assumed to simplify
the numerical calculation.

The diquark-antidiquark picture of tetraquark states has been
discussed by several authors~\cite{jaffe3,Maiani,black,jaffe4}. In
the present work, the scalar mesons below 1 GeV are studied as
diquark-antidiquark systems in the flux-tube model. Two colour
structures for a tetraquark state are shown in Fig.1, where the
solid dot represents a quark while the hollow dot represents an
anti-quark, $\mathbf{r}_i$ is quark's position, $\mathbf{y}_i$
represents a junction where three strings (flux tubes) meet. A
thin line connecting a quark and a junction (an antiquark)
represents a fundamental representation, {\em i.e.} colour
triplet. A thick line connecting two junctions is for a colour
sextet or other representations, namely a compound string. The
different types of string may have different
stiffness~\cite{mit,semay,bali}. In Fig.1 b, colour couplings
satisfying overall colour singlet of tetra-quark are $\left [
[qq]_{\bar{\mathbf{3}}}[\bar{q}\bar{q}]_{\mathbf{3}}\right
]_{\mathbf{1}}$ and $\left [
[qq]_{\mathbf{6}}[\bar{q}\bar{q}]_{\bar{\mathbf{6}}}\right
]_{\mathbf{1}}$, the subscripts represent the dimensions of colour
representations.

In the flux-tube model with quadratic confinement potential, which
is believed to be flavour independent, of the tetraquark state
with a diquark-antidiquark structure has the following
form~\cite{deng},
\begin{eqnarray}
V^{CH}&=&k\left[ (\mathbf{r}_1-\mathbf{y}_1)^2
+(\mathbf{r}_2-\mathbf{y}_1)^2+(\mathbf{r}_3-\mathbf{y}_2)^2\right. \nonumber \\
&+&
\left.(\mathbf{r}_4-\mathbf{y}_2)^2+\kappa_d(\mathbf{y}_1-\mathbf{y}_2)^2\right],
\end{eqnarray}
where $k$ is the stiffness of the string with the
fundamental representation \textbf{3} which is determined by meson
spectrum, $k\kappa_d$ is the compound string stiffness. The
compound string stiffness parameter $\kappa_d$~\cite{bali} depends
on the colour representation, $\mathbf{d}$, of the string,
\begin{equation}
 \kappa_{\mathbf{d}}=\frac{C_\mathbf{d}}{C_\mathbf{3}},
\end{equation}
where $C_\mathbf{d}$ is the eigenvalue of the Casimir operator
associated with the $SU(3)$ colour representation $\mathbf{d}$ of
the string. $C_\mathbf{3}=\frac{4}{3}$,
$C_\mathbf{6}=\frac{10}{3}$ and $C_\mathbf{8}=3$.
\begin{figure}
\includegraphics[width=7cm]{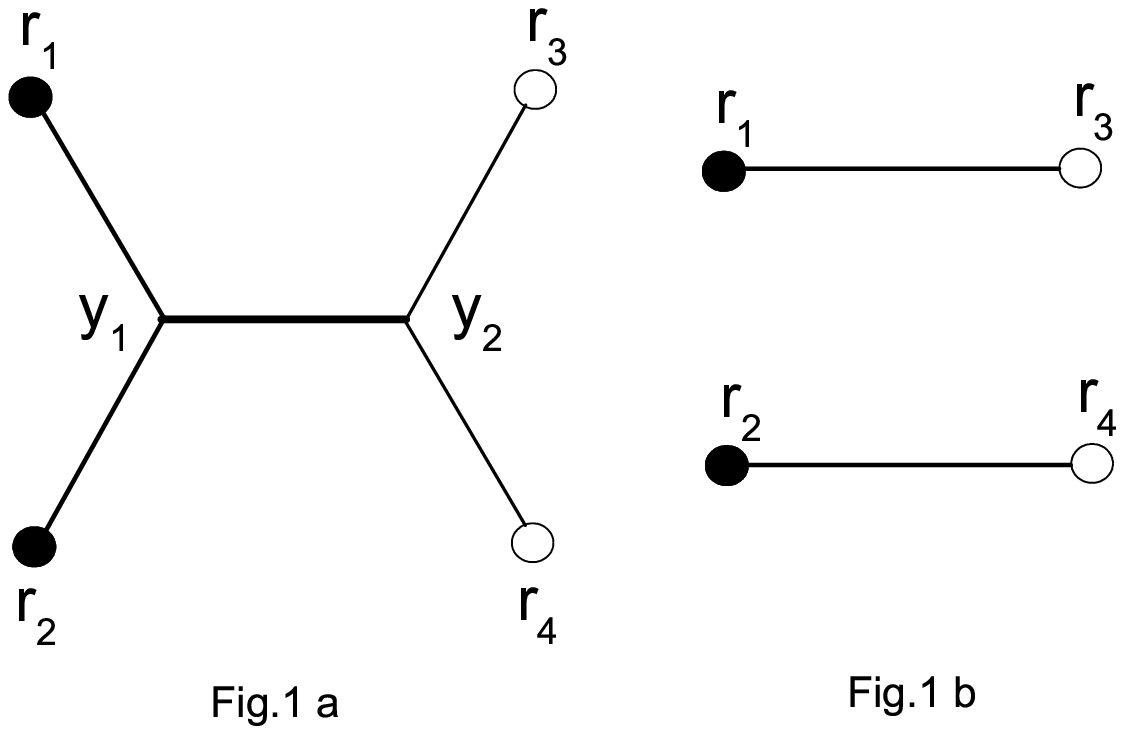}
\caption{Two colour structures for a tetraquark
state.\label{fig1}}
\end{figure}

For given quark positions $\mathbf{r}_i~(i=1,...,4)$, we can fix
the positions of the junctions $\mathbf{y}_i~(i=1,2)$ by
minimizing the energy of the system. After fixing $\mathbf{y}_i$,
a set of canonical coordinates $\mathbf{R}_i~(i=1, ... ,4)$ is
introduced to simplify the expressions of the potential, which are
read as,
\begin{eqnarray}
\mathbf{R}_{1}&=&\sqrt{\frac{1}{2}}(\mathbf{r}_1-\mathbf{r}_2),~~
\mathbf{R}_{2}=\sqrt{\frac{1}{2}}(\mathbf{r}_3-\mathbf{r}_4), \nonumber \\
\mathbf{R}_{3}&=&\sqrt{\frac{1}{4}}(\mathbf{r}_1+\mathbf{r}_2-\mathbf{r}_3-\mathbf{r}_4),\\
\mathbf{R}_{4}&=&\sqrt{\frac{1}{4}}(\mathbf{r}_1+\mathbf{r}_2+\mathbf{r}_3+\mathbf{r}_4).
\nonumber
\end{eqnarray}
Therefore, the minimum of the confinement interaction has the
following form,
\begin{eqnarray}
V^{CH}_{\mbox{min}} & = &
k\left(\mathbf{R}_{1}^2+\mathbf{R}_{2}^2+\frac{\kappa_d}{1+\kappa_d}\mathbf{R}_{3}^2
\right)
\end{eqnarray}
Taking into account the potential energy shift, the confinement
potential $V^C_{min}$ used in the present calculation has the
following form
\begin{eqnarray}
V^{CH}_{\mbox{min}} =
k\left[(\mathbf{R}_1^2-\Delta)+(\mathbf{R}_2^2-\Delta)+
\frac{\kappa_d}{1+\kappa_d}(\mathbf{R}_3^2-\Delta)\right]
\end{eqnarray}
Carlson and Pandharipande also considered similar flux tube energy
shift which they assumed to be proportional to the number of
quarks $N$~\cite{carlson}. Obviously, the confinement potential
$V^C$ is a multi-body interaction rather than a two-body
interaction. It should be emphasized here that our approach is
different from that in Iwasaki's work~\cite{Iwasaki}, where the
four-body problem is simplified to two-body one by treating
diquark as an antiquark and antidiquark as a quark.

With regard to the mesons $f_0(980)$ and $a_0(980)$, they are also
interpreted as $K\bar{K}$ molecular states with $I=0$ and $I=1$,
respectively~\cite{weinstein0,giacosa,branz1,branz2,Janssen,Kaminski,Oller1,Oller2}.
In the flux tube model, the confinement potential of $K\bar{K}$
molecular states can be written as
\begin{eqnarray}
V^{CM}_{\mbox{min}} =
k\left[((\mathbf{r}_1-\mathbf{r}_3)^2-\Delta)+((\mathbf{r}_2-\mathbf{r}_4)^2-\Delta)\right]
\end{eqnarray}
where $q_1$ ($q_2$) and $\bar{q}_3$ ($\bar{q}_4$) compose a $K$
($\bar{K}$) meson, see Fig.1 b. In fact, the mesons $f_0(980)$ and
$a_0(980)$, if they are really tetraquark systems, should be the
superposition of the diaquark-antidiquark state and $K\bar{K}$
molecular state. When two mesons $K$ and $\bar{K}$ are separated
largely, the dominant component of the system should be two
isolated color singlet mesons because other hidden color flux tube
structures are suppressed due to a confinement. With the
separation reduction, a loose $K\bar{K}$ molecular state may be
formed if the attractive force between $K\bar{K}$ is strong
enough. Especially when they are close enough to be within the
range of a confinement (about 1 fm), the diquark-antidiquark state
and the $K\bar{K}$ molecular state may appear due to the
excitation and rearrangements of flux tubes and junctions. In this
case, the confinement potential of a tetraquark system
$ns\bar{n}\bar{s}$ should be taken to be the minimum of two flux
tube structures. It reads
\begin{eqnarray}
V^C_{\mbox{min}} = \mbox{min} \left[ V^{CM}_{\mbox{min}},
V^{CH}_{\mbox{min}} \right].
\end{eqnarray}

The other parts of the Hamiltonian are rest masses, kinetic
energies, one-gluon-exchange potential and
Goldstone-boson-exchange potentials~\cite{deng},
\begin{widetext}
\begin{eqnarray*}
H&=&\sum_{i=1}^4 \left( m_i+\frac{\mathbf{p}_i^2}{2m_i}
\right)-T_{CM} + V^C + \sum_{i>j}^4 (V_{ij}^G+V_{ij}^B),
\\
V_{ij}^B&=&
 v^{\pi}_{ij} \sum_{a=1}^3 F_i^a F_j^a
+v^{K}_{ij} \sum_{a=4}^7 F_i^a F_j^a +v^{\eta}_{ij} (F^8_i\cdot
F^8_j\cos \theta_P-F^0_i\cdot F^0_j\sin \theta_P),  \\
v^{\chi}_{ij}&=& \frac{g^2_{ch}}{4\pi}\frac{m^3_{\chi}}{12m_im_j}
\frac{\Lambda^{2}_{\chi}}{\Lambda^{2}_{\chi}-m_{\chi}^2}
 \boldmath{\mbox{$\sigma$}}_i\cdot
 \boldmath{\mbox{$\sigma$}}_j\left[ Y(m_\chi r_{ij})-
\frac{\Lambda^{3}_{\chi}}{m_{\chi}^3}Y(\Lambda_{\chi} r_{ij})
\right],~~~~ \chi=\pi,K,\eta. \\
V_{ij}^G&=&{\frac{\alpha _{s}}{4}}\mathbf{\lambda}^{c}%
_{i}\cdot\mathbf{\lambda}^{c}_{j}\left[{\frac{1}{r_{ij}}}-{\frac{\pi}{2}}
\delta(\mathbf{r}_{ij}) \left(
{\frac{1}{m_i^2}}+{\frac{1}{m_j^2}}+{\frac{4}{3m_im_j}}
 \boldmath{\mbox{$\sigma$}}_i\cdot
 \boldmath{\mbox{$\sigma$}}_j \right) \right]
\end{eqnarray*}
\end{widetext}
where $T_{CM}$ is the center-of-mass kinetic energy,
$\mathbf{F}_i,\lambda_i$ are flavour, colour SU$_3$ Gell-mann
matrices. $Y(x)$ is the standard Yukawa function and all other
symbols have their usual meanings. The $\delta$-function should be
regularized~\cite{PRL44,salamanca}
\begin{equation}
\delta(\mathbf{r}_{ij})={\frac{1}{4\pi}}{\frac{e^{-r_{ij}/r_0(\mu)}}
{r_{ij}r_0^2(\mu)}}
\end{equation}
where $\mu$ is the reduced mass of $q_i$ and $q_j$ and
$r_0(\mu)=\hat{r}_0/\mu$. The effective scale-dependent strong
coupling constant is given by~\cite{salamanca}
\begin{equation}
\alpha_s(\mu)=\frac{\alpha_0}
{\ln\left[\frac{\mu^{2}+\mu_0^2}{\Lambda_0^2}\right]}.
\end{equation}

\section{Wave functions and gaussian expansion method}

The total wave function of a diquark-antidiquark state can be
written as a sum of the following direct products of colour,
isospin and spatial-spin terms,
\begin{eqnarray}
\Phi^{[qq][\bar{q}\bar{q}]}_{IJ_TM_T}&=&\sum_{l,s,c,I}\xi_{l,s,c,I}
\left[ \left[
\left[\phi^G_{l_1m_1}(\mathbf{r})\eta_{s_1m_{s_1}}\right]_{J_1M_1}\right. \right. \nonumber \\
&\times& \left. \left.
\left[\psi^G_{l_2m_2}(\mathbf{R})\eta_{s_2m_{s_2}}\right]_{J_2M_2}
\right ]_{J_{12}M_{12}}\right. \nonumber \\
&\times& \left. \chi^G_{LM}(\mathbf{X}) \right]_{J_TM_T}
\left[\eta_{i_1m_{i_1}}\eta_{i_2m_{i_2}}\right]_I\nonumber \\
&\times&\left[\chi_{c_1w_{1}}\chi_{c_2w_{2}}\right]_1,
\end{eqnarray}
Here $I$ and $J_T$ are the total isospin and total angular
momentum respectively. $\eta_{s_1m_{s_1}}$ ($\eta_{s_2m_{s_2}}$),
$\eta_{i_1m_{i_1}}$ ($\eta_{i_2m_{i_2}}$) and $\chi_{c_1w_{1}}$
($\chi_{c_2w_{2}}$) are the spin, flavour and colour wave
functions of, diquark (anti-diquark), respectively. [~]'s denote
Clebsh-Gordan coefficients coupling. The coefficient
$\xi^{IJ_T}_{l,s,i,c,L}$ is determined by diagonalizing the
Hamiltonian, subscripts $l,s,i,c,L$ represent all possible
intermediate quantum numbers, therefore our calculations are
multi-channel coupling calculations. The Jacobi coordinates of the
tetraquark are defined as
\begin{eqnarray}
\mathbf{r}&=&\mathbf{r}_1-\mathbf{r}_2,\ {} \ {}
\mathbf{R}=\mathbf{r}_3-\mathbf{r}_4\nonumber\\
\mathbf{X}&=&\frac{m_1\mathbf{r}_1+m_2\mathbf{r}_2}{m_1+m_2}
-\frac{m_3\mathbf{r}_3+m_4\mathbf{r}_4}{m_3+m_4}\\
\mathbf{R_{CM}}&=&\frac{m_1\mathbf{r}_1+m_2\mathbf{r}_2
+m_3\mathbf{r}_3+m_4\mathbf{r}_4}{m_1+m_2+m_3+m_4}\nonumber
\end{eqnarray}
where Particles 1 and 2 are two quarks and Particle 3 and 4 are
two anti-quarks. $L$, $l_1$ and $l_2$ are the orbital angular
momenta associated with the coordinates of $\mathbf{X}$,
$\mathbf{r}$ and $\mathbf{R}$, respectively. The calculation is
done in the center-of-mass coordinate system
($\mathbf{R_{CM}}=0$). The tetra-quark state is an overall colour
singlet with well defined parity $P=(-1)^{l_1+l_2+L}$, isospin $I$
and the total angular momentum $J_T$. For scalar mesons, we set
the angular momentum $L$, $l_1$ and $l_2$ to be zero.

For the colour part, the colour singlet is constructed in the
following two ways, $\chi_c^1=\bar{3}_{12}\otimes3_{34}$,
$\chi_c^2=6_{12}\otimes\bar{6}_{34}$, both ``good" diquark and
``bad" diquark are included. With respect to the flavour part, the
flavour wave function reads as $\eta_I=\eta_{12}\otimes\eta_{34}$.
Taking into account all degrees of freedom, the Pauli principle
must be satisfied for each subsystem of the identical quarks or
anti-quarks. To obtain a reliable solution of few-body problem, a
high precision method is indispensable. In this work, the
GEM~\cite{GEM}, which has been proven to be rather powerful to
solve few-body problem, is used to do the calculations. In GEM,
three relative motion wave functions are written as,
\begin{eqnarray}
\phi^G_{l_1m_1}(\mathbf{r})&=&\sum_{n_1=1}^{n_{1max}}c_{n_1}N_{n_1l_1}r^{l_1}
e^{-\nu_{n_1}r^2}Y_{l_1m_1}(\hat{\mathbf{r}})\nonumber\\
\psi^G_{l_2m_2}(\mathbf{R})&=&\sum_{n_2=1}^{n_{2max}}c_{n_2}N_{n_2l_2}R^{l_2}
e^{-\nu_{n_2}R^2}Y_{l_2m_2}(\hat{\mathbf{R}})\nonumber\\
\chi^G_{LM}(\mathbf{X})&=&\sum_{n_3=1}^{n_{3max}}
c_{n_3}N_{LM}X^{L}e^{-\nu_{n_3}X^2}Y_{LM}(\hat{\mathbf{X}})\nonumber
\end{eqnarray}
The Gaussian size parameters are taken as the following geometric
progression numbers
\begin{eqnarray}
\nu_{n}=\frac{1}{r^2_n},& r_n=r_1a^{n-1},&
a=\left(\frac{r_{n_{max}}}{r_1}\right)^{\frac{1}{n_{max}-1}}.
\end{eqnarray}

Within the framework of the flux tube model, the wavefunctions of
a $K\bar{K}$ molecular state can be expressed as
\begin{eqnarray}
\Phi^{K\bar{K}}_{IJ_TM_T}&=&\sum_{M,S,I}\xi_{M,S,I}\left [\left [
\phi^G_{K}(\mathbf{r})\psi^G_{\bar{K}}(\mathbf{R})\chi^G_{LM}(\mathbf{X})\right
] \right. \nonumber \\
&\times& \left.\eta_S\right ]_{J_TM_T} \eta_I\chi_c
\end{eqnarray}
The details of the wavefunctions are omitted and similar to those
of a diquark-antiquark state.

\section{Numerical results and discussions}

Now we turn to the calculation on tetraquark states with diquark
antiquark structures. The model parameters are fixed by
reproducing the ordinary meson spectrum and are listed in Table
\ref{tab1}. the meson spectrum can be reproduced very well.
Because the flux-tube model is reduced to the ordinary quark model
for a $q\bar{q}$ system, the obtained meson spectra (from light to
heavy) are similar to those of other work, e.g.
Ref.~\cite{salamanca}. Parts of the calculated meson spectra are
shown in Table \ref{tab2}. The experimental values are taken from
PDG compilation~\cite{PDG}.

\begin{table}
\caption{The model parameters (Set I). The masses of
$\pi,K,\eta$ take the experimental values.\label{tab1}}
\footnotesize
\begin{tabular*}{80mm}{c@{\extracolsep{\fill}}cccccc}
\hline
$m$ & $m_s$ & $k$ & $\hat{r}_0$ & $\Lambda_0$ & $\mu_0$ \\
MeV & MeV & MeV fm$^{-2}$ & MeV fm & fm$^{-1}$ &
fm$^{-1}$  \\
313 & 520 & 213.3 & 30.85 & 0.187 & 0.113 \\ \hline $\Delta$ &
$\alpha_0$ & $\Lambda_{\pi}$ &
 $\Lambda_K=\Lambda_{\eta}$ & $\theta_P$ \\
fm$^2$ & - & fm$^{-1}$ & fm$^{-1}$ & - \\
 0.5 & 4.25 & 4.2 & 5.2 & 15$^{o}$ \\ \hline
\end{tabular*}
\end{table}

\begin{table}
\caption{The meson spectra (unit: MeV). \label{tab2}}
\footnotesize
\begin{tabular*}{80mm}{c@{\extracolsep{\fill}}cccccc}
\toprule
Meson & $\pi$ & K & $\rho$ & $K^*$ & $\omega$ & $\phi$ \\
 Cal. & 139 & 502 & 761 & 897 & 735 & 1023 \\
 Exp. & 139 & 496 & 770 & 898 & 780 & 1020 \\
$\sqrt{\langle r^2\rangle}$(fm) & 0.57 & 0.60
 & 1.05 & 0.96 & 1.02 & 0.85  \\ \bottomrule
\end{tabular*}
\end{table}

The energies of scalar meson states can be obtained by solving the
four-body Schr\"{o}dinger equation
\begin{eqnarray}
(H-E)\Phi_{IJ_TM_T}=0
\end{eqnarray}
with Rayleigh-Ritz variational principle. In GEM the calculated
results are converged with $n_{1max}$=6, $n_{2max}=6$ and
$n_{3max}=6$. The minimum and maximum ranges of the bases are 0.1
fm and 2.0 fm for coordinates $\mathbf{r}$, $\mathbf{R}$ and
$\mathbf{X}$, respectively.

Quark contents and the corresponding masses in the three different
quark models for the light scalar mesons as tetra-quark states are
shown in Table \ref{tab3}, where $n$ stands for a non-strange
quark ($u$ or $d$) while $s$ stands for a strange quark, $I$ and
$N$ denote the total isospin and principal quantum number of the total
radial excitation, $S$, $L$ and $J$ have their usual meanings.
``Naive" stands for the naive quark model, where only
one-gluon-exchange potential is taken into account in addition to
the additive two-body confinement~\cite{Isgur}. ``Chiral" stands
for the chiral quark model, where one-gluon-exchange and
one-Goldstone-boson-exchange are included besides the additive
two-body confinement~\cite{salamanca}. The masses in the naive and
chiral model are much higher (several hundreds MeV) than those in
the flux-tube model, the origin of this discrepancy mainly comes
from the different type of confinement interaction, a two-body
confinement potential is applied in the naive and chiral model,
whereas a multi-body interaction confinement is used in the
flux-tube mode. Zou {\em et al.} studied scalar mesons in the
quark model by introducing three-body confinement interaction.
Their study also indicates that the multi-body confinement
potential, instead of two-body interaction, should be applied in
the study of multi-quark states~\cite{Zou}. The naive quark model
gives the highest masses, due to the absence of Goldstone boson
exchange, which induces additional attraction for the tetraquark
system.

\begin{table}
\caption{Numerical results for three models (unit: MeV).
\label{tab3}} \footnotesize
\begin{tabular*}{80mm}{c@{\extracolsep{\fill}}cccccc}
\hline Flavour & $nn\bar{n}\bar{n}$ & $nn\bar{n}\bar{n}$ &
  $nn\bar{n}\bar{n}$ \\
 $IJ^{P}$ & $00^+$ & $00^+$ & $10^+$ \\
 $N^{2S+1}L_J$ & $0^1S_0$ & $1^1S_0$ & $0^1S_0$ \\
 \hline
 Naive & 938 & 1431 & 1431 \\
 Chiral & 666 & 1237 & 1406 \\
 Flux-tube & 587 & 1019 & 1210 & \\ \hline
 Candidate & $\sigma$ & $f_0(980)$? & --- & \\
 Mass & $541\pm 39$~\cite{BESsigma}  & $980\pm 10$~\cite{PDG}  & \\ \hline
 Flavour & $nn\bar{n}\bar{s} $ &$ns\bar{n}\bar{s}$ &
 $ns\bar{n}\bar{s}$ \\
 $IJ^{P}$ & $\frac{1}{2}0^+$ & $00^+$ & $10^+$ \\
 $N^{2S+1}L_J$ & $0^1S_0$ & $0^1S_0$ & $0^1S_0$ \\
 Naive & 1216 & 1456 & 1456 \\
 Chiral & 1122 & 1454 & 1454 \\
 Flux-tube & 948 & 1314 & 1318 \\ \hline
 Candidate & $\kappa$ & --- & --- \\
 Mass & $826\pm 49^{+49}_{-34}$~\cite{bes2} & &  \\ \hline
\end{tabular*}
\end{table}

In the framework of the flux tube model, it can be seen from Table
\ref{tab3} that the lowest masses of $nn\bar{n}\bar{n}$ and
$nn\bar{n}\bar{s}$ system are 587 MeV and 948 MeV, which are close
to the masses of the mesons $\sigma$ and $\kappa$. If the
existence of the mesons $\sigma$ and $\kappa$ is further
confirmed, the tetraquark state is a possible interpretation. This
interpretation is in agreement with many other
studies~\cite{jaffe1,jaffe2,alford,maiani,Pelaez1,Pelaez2,thooft}.
Prelovsek {\em et al.} recently studied the light scalar mesons
$\sigma$ and $\kappa$ by lattice QCD simulation, they also found
that $\sigma$ and $\kappa$ have sizable tetra-quark components
$nn\bar{n}\bar{n}$ and $nn\bar{n}\bar{s}$,
respectively~\cite{Prelovsek}. In order to check the dependence of
numerical results on the model parameters, we make the same
calculations of scalar mesons with another set of parameters which
are listed in Table \ref{tab4} (the unchanged parameters are not
listed). The almost same meson spectrum is obtained. The results
for tetraquark states are shown in Table \ref{tab5}. Comparing
Table \ref{tab4} and \ref{tab5}, our results are rather stable
against the variation of model parameters.

\begin{table}
\caption{The model parameters (Set II). \label{tab4}}
\footnotesize
\begin{tabular*}{80mm}{c@{\extracolsep{\fill}}cccccc}
\hline
$k$ & $\hat{r}_0$ & $\Delta$ & $\alpha_0$ \\
 MeV fm$^{-2}$ & MeV fm & fm$^2$ & - \\ \hline
 267 & 30.0 & 0.6 & 4.09   \\ \hline
\end{tabular*}
\end{table}

\begin{table}
\caption{Numerical results in the flux-tube model (unit: MeV).
\label{tab5}} \footnotesize
\begin{tabular*}{80mm}{c@{\extracolsep{\fill}}cccccc}
\hline
 Flavour&$nn\bar{n}\bar{n}$&$nn\bar{n}\bar{n}$&$nn\bar{n}\bar{n}$
 &$nn\bar{n}\bar{s}$&$ns\bar{n}\bar{s}$&$ns\bar{n}\bar{s}$\\
 $IJ^{P}$&$00^+$&$00^+$&$10^+$&$\frac{1}{2}0^+$&$00^+$&$10^+$\\
 $N^{2S+1}L_J$&$0^1S_0$&$1^1S_0$&$0^1S_0$&$0^1S_0$&$0^1S_0$&$0^1S_0$\\
 \hline
 Flux-tube&531&969&1180&908&1270&1275  \\ \hline
\end{tabular*}
\end{table}

The mesons $\sigma$ and $\kappa$, if they have diquark antiduqark
structures, can not decay into two colourful hadrons directly due
to a colour confinement. They must transform into colour singlet
mesons by means of breaking and rejoining flux tubes before
decaying into colour singlet mesons. This decay mechanism is
similar to the compound nucleus formation and therefore should
induce a resonance which is named a ``colour confined, multi-quark
resonance" state~\cite{ping,resonance}. The large decay width of
the mesons $\sigma$ and $\kappa$ may be qualitatively explained if
the arrangement and rupture of the flux tubes are fast enough, the
systematic investigation of the decay is left for the future work.

In the case of the mesons $f_0(980)$ and $a_0(980)$, many
theoretical studies assumed them as tetra-quark states with quark
content $ns\bar{n}\bar{s}$ with isospin $I=0$ and $I=1$,
respectively. The masses for the tetraquark states
$ns\bar{n}\bar{s}$ are much higher, about 300 MeV, than the
experimental values even in the flux tube model, see Table 3 and
Table 5. Therefore, their main components seem to be not the
tetraquark state $ns\bar{n}\bar{s}$ in the quark models. Instead,
the mass of the first radial excited state of the state
$nn\bar{n}\bar{n}$ is 1019 MeV, which is close to the mass of the
meson $f_0(980)$. Taking the meson $f_0(980)$ as
$nn\bar{n}\bar{n}$ state is consistent with Vijande's work on the
nature of scalar mesons~\cite{Vijande}. The observed
$f_0(980)\rightarrow K\bar{K}$ process can be explained by the
mixing of $nn\bar{n}\bar{n}$, $ns\bar{n}\bar{s}$ and $s\bar{s}$
{\em et al.} strange quark components~\cite{Vijande}. This work is
being done in our group. Pel\'{a}ez also suggested that three
scalar mesons, $\sigma$, $\kappa$ and $f_0(980)$, have dominant
tetraquark component, whereas the meson $a_0(980)$ might be a more
complicated system~\cite{Pelaez}, which is also consistent with
our results.

The other well known interpretations of the mesons $f_0(980)$ and
$a_0(980)$ are the $K\bar{K}$ bound states with isospin $I=0$ and
$I=1$, respectively, because the experimental values are very
close to the threshold of two mesons $K$ and $\bar{K}$. Within the
quark models, the interactions between two quarks related to the
color and spin factors,
$\mathbf{\lambda}_i\cdot\mathbf{\lambda}_j$ and
$\mathbf{\sigma}_i\cdot\mathbf{\sigma}_j$, which are zero between
the mesons $K$ and $\bar{K}$, therefore the $K\bar{K}$ bound state
is hard to be formed. The coupling calculations on the diquark
antidiquark state and the $K\bar{K}$ state indicate that the
$K\bar{K}$ bound state still can not be formed. The arguments are:
(i) the interactions between $K$ and $\bar{K}$ are equal to zero,
the coupling interaction between the diquark antidiquark state and
the $K\bar{K}$ state is weak; and (ii) the relative kinetic energy
between two mesons $K$ and $\bar{K}$ is not small due to the small
mass of the $K$ ($\bar{K}$) meson. The two factors are not
beneficial to form a bound state.

\section{Summary}
The comparative studies of the three quark models on the light
scalar mesons indicate that a multibody confinement potential,
instead of a two-body confinement potential proportional to a
colour factor, plays an important role in a multiquark state,
which can reduce the energy of a multiquark state because it
avoids the appearance of the anti-confinement in color symmetrical
quark (antiquark) pairs.

In the flux-tube model, the mesons $\sigma$ and $\kappa$ can be
assigned as diquark antidiquark states $nn\bar{n}\bar{n}$ and
$nn\bar{n}\bar{s}$ with $J^P=0^+$, respectively, which can be
named as ``colour confined, multi-quark resonance" states. The
interpretation of the mesons $f_0(980)$ and $a_0(980)$ as the
tetraquark states $ns\bar{n}\bar{s}$ with $I=0$ and $I=1$,
respectively, would give a much higher mass (about 300 MeV) than
the experimental data. The studies on the mixing of the diquark
antidiquark state $ns\bar{n}\bar{s}$ and the $K\bar{K}$ state
indicate that the $K\bar{K}$ molecular state does not exist in the
quark models due to the weak coupling and a big relative kinetic
energy between the mesons $K$ and $\bar{K}$. However, in our
calculation the mass of the first radial excitation of the diquark
antidiquark state $nn\bar{n}\bar{n}$ is close to the mass of the
meson $f_0(980)$. The problem of this assignment, the small decay
width of the meson $f_0(980)\rightarrow K\bar{K}$, can be
accounted for by the mixing of $ns\bar{n}\bar{s}$ and $s\bar{s}$
{\em et al.} strange quark components with the $nn\bar{n}\bar{n}$
state.

At present the nature of the scalar mesons is still an open
question, the interpretation of the scalar mesons as tetraquark
states is a possibility. In fact, the scalar mesons should be the
superpositions of $q\bar{q}$, $qq\bar{q}\bar{q}$ and other
components in a Fock space expansion approach and the dominant one
is determined by quark dynamics. The mixing between two-body and
four-body configurations would require the knowledge of the
quark-antiquark pair creation-annihilation interaction, which is
being calculated in our group by using a $^3P_0$ model tentatively.\\


\begin{thebibliography}{99}
\bibitem{bes2}  Ablikim M, {\em et al.} [BES Collaboration], Phys. Lett. B, 2011, {\bf 698}, 183-190
\bibitem{neutralk}Ablikim M, {\em et al.} [BES Collaboration], Phys. Lett. B, 2006, {\bf 633}, 681-690
\bibitem{jaffe3} Jaffe R L, Phys. Rep., 2005, {\bf 409}, 1-45
\bibitem{Amsler} Amsler C and T\"{o}rnqvist N A, Phys. Rep., 2004, {\bf 389}, 61-118
\bibitem{jaffe1} Jaffe R L, Phys. Rev. D, 1977, {\bf 15}, 267-280
\bibitem{jaffe2} Jaffe R L, Phys. Rev. D, 1977, {\bf 15}, 281-289
\bibitem{alford} Alford M G and Jaffe R L, Nucl. Phys. B, 2000, {\bf 578}, 367-382
\bibitem{maiani} Maiani L, Piccinini F, Polosa A D and Riquer V, Phys. Rev. Lett., 2004, \textbf{93}, 212002: 1-4
\bibitem{Pelaez1} Pelaez J R, Phys. Rev. Lett., 2004, {\bf 92}, 102001:1-4
\bibitem{Pelaez2} Pelaez J R and Rios G, Phys. Rev. Lett., 2006, {\bf 97}, 242002: 1-4
\bibitem{thooft} Hooft G.'t, Isidori G, Maiani L, Polosa A D and Riquer V, Phys. Lett. B, 2008, {\bf 662}, 424-430
\bibitem{weinstein0} Weinstein J and Isgur N, Phys. Rev. Lett., 1982, {\bf 48}, 659-662
\bibitem{giacosa} Giacosa F, Gutsche T and Lyubovitskij V E, Phys. Rev. D, 2008, {\bf 77}, 034007: 1-6
\bibitem{branz1} Branz T, Gutsche T and Lyubovitskij V E, Eur. Phys. J. A, 2008, {\bf 37}, 303-317
\bibitem{branz2} Branz T, Gutsche T and Lyubovitskij V E, Phys. Rev. D, 2008, {\bf 78}, 114004: 1-10
\bibitem{Janssen} Janssen G, Pearce B C, Holinde K and J. Speth, Phys. Rev. D, 1995, {\bf 52}, 2690-2700
\bibitem{Kaminski} Kaminski R, Lesniak L and  Maillet J P, Phys. Rev. D, 1994, {\bf 50}, 3145-3157
\bibitem{Oller1} Oller J A and Oset E, Nucl. Phys. A, 1997, {\bf 620}, 438-456
\bibitem{Oller2} Oller J A and Oset E, Phys. Rev. D, 1999, {\bf 60}, 074023: 1-22
\bibitem{van} Van Beveren E and Rupp G, Eur. Phys. J. C, 2001, {\bf 22}, 493-501
\bibitem{celenza} Celenza L S, Gao S F, Huang B, Wang H and Shakin C M, Phys. Rev. C, 2000, {\bf 61}, 035201: 1-17
\bibitem{Umekawa} Umekawa T, Naito K, Oka M and Takizawa M, Phys. Rev. C, 2004, {\bf 70}, 055205: 1-9
\bibitem{Dai} Dai Y B and Wu Y L, Eur. Phys. J. C, 2005, {\bf 39}, S1-S8
\bibitem{Bhavyashri} Bhavyashri, Vijaya Kumar K B, Ma Y.L and Prakash A, Int. J. Mod. Phys. A, 2009, {\bf 24}: 4209-4220
\bibitem{Vijande} Vijande J, Valcarce A, Fernandez F and Silvestre-Brac B, Phys. Rev. D, 2005, {\bf 72}, 034025: 1-8
\bibitem{Napsuciale} Napsuciale M and Rodr\'{i}guez S, Phys. Lett. B, 2004, {\bf 603}, 195-2002
\bibitem{Dmitrasinovic} Dmitrasinovic V, Phys. Rev. D, 2003, {\bf 67}, 114007: 1-12
\bibitem{Pepin} Pepin S and Stancu F, Phys. Rev. D, 2002, {\bf 65}, 054032: 1-10
\bibitem{deng} Deng C R, Ping J L, Wang F and Goldman T, Phys. Rev. D, 2010, {\bf 82}, 074001: 1-8
\bibitem{ping} Ping J L, Deng C R, F. Wang and Goldman T, Phys. Lett. B, 2008, {\bf 659}, 607-611
\bibitem{GEM} Hiyama E, Kino Y and Kamimura M, Prog. Part. Nucl. Phys., 2003, {\bf 51}, 223-307
\bibitem{ww} Wang F and Wong C W, Nuovo Cimento A, 1985, {\bf 86}, 283-289
\bibitem{oka1} Oka M and Horowitz C J, Phys. Rev. D, 1985, {\bf 31}, 2773-2779
\bibitem{oka2} Oka M, Phys. Rev. D, 1985, {\bf 31}, 2274-2287
\bibitem{lattice} Alexandrou C, de Forcrand P and Tsapalis A, Phys. Rev. D, 2002, {\bf 65}, 054503: 1-7
\bibitem{lattice0} Takahashi T T, Suganuma H, Nemoto Y and H. Matsufuru, Phys. Rev. D, 2002, {\bf 65}, 114509: 1-19
\bibitem{latt1} Okiharu F, Suganuma H and Takahashi T T, Phys. Rev. Lett., 2005, {\bf 94}, 192001: 1-4
\bibitem{Maiani} Maiani L, Piccinini F, Polosa A D and Riquer V, Phys. Rev. Lett., 2004, {\bf 93}, 212002: 1-4
\bibitem{black} Black D, Fariborz A and Schechter J, Phys. Rev. D, 2000, {\bf 61}, 074001: 1-10
\bibitem{jaffe4} Jaffe R L and Wilczek F, Phys. Rev. Lett., 2003, {\bf 91}, 232003: 1-4
\bibitem{mit} Johnson K and Thorn C B, Phys. Rev. D, 1976, {\bf 13}, 1934-1919
\bibitem{semay} Semay C, Eur. Phys. J. A, 2004, {\bf 22}, 353-354
\bibitem{bali} Bali G S, Phys. Rev. D, 2000, {\bf 62}, 114503: 1-11
\bibitem{carlson} Carlson J and Pandharipande V R, Phys. Rev. D, 1991, {\bf 43}, 1652-1658
\bibitem{Iwasaki} Iwasaki M and Fukutome T, Phys. Rev. D, 2005, {\bf 72}, 094016: 1-5
\bibitem{PRL44} Bhaduri R K, Cohler L E and Nogami Y, Phys. Rev. Lett., 1980, {\bf 44}, 1369-1372
\bibitem{salamanca} Vijande J, Fernandez F and Valcarce A, J. Phys. G, 2005, \textbf{31}, 481:
\bibitem{PDG} Nakamura K, {\em et al.} [Particle Data Group], J. Phys. G, 2010, {\bf 37}, 075021:
\bibitem{Isgur} Isgur N, in {\em The New Aspects of Subnuclear Physics}, edited by A. Zichichi (Plenum, New York, 1980), p.107
\bibitem{Zou} Zou F Y, Chen X L and W.Z. Deng, Chin. Phys. C, 2008, {\bf 32}: 515-520
\bibitem{Prelovsek} Prelovsek S, Draper T and C.B. Lang {\em et al.}, Phys. Rev. D, 2010, {\bf 82}, 094507: 1-18
\bibitem{resonance} Wang F, Ping J L, Pang H R and L Z Chen, Nucl. Phys. A, 2007, {\bf 790}, 493c-497c
\bibitem{Pelaez} Pel\'{a}ez J R, Phys. Rev. Lett., 2005, {\bf 92}, 102001: 1-4

\end{thebibliography}
\end{document}